\documentclass[12pt]{article}
\usepackage{amsmath,amsthm,amsfonts}

\input epsf.tex
\newdimen\epsfxsize
\newdimen\epsfysize
\def\qed{\vrule height5pt width3pt depth.5pt}

\theoremstyle{plain}
\newtheorem{thm}{Theorem}[section]

\newtheorem{prop}[thm]{Proposition}

\newtheorem{rem}{Remark}[section]

\begin{document}

\title{Unitary Solutions to the Yang-Baxter Equation in Dimension Four}

\author{H. A.  Dye \\
Department of Mathematics, Statistics, and Computer Science  \\
  University of Illinois at Chicago \\
  851 South Morgan St \\
 Chicago, IL  60607-7045 \\ 
 dye@math.uic.edu }

\maketitle

\begin{abstract}
In this paper, we determine 
 all unitary solutions to the Yang-Baxter equation in 
dimension four. Quantum computation motivates this study.
 This set of solutions will assist in clarifying the
relationship between quantum entanglement and topological entanglement.
 We present a variety of facts about the Yang-Baxter
equation for the reader unfamiliar with the equation.
\end{abstract}
\noindent {\bf Acknowledgement.} This effort was sponsored by the
Defense
Advanced Research Projects Agency (DARPA) and Air Force Research
Laboratory, Air
Force Materiel Command, USAF, under agreement F30602-01-0522.
 The U.S. Government is authorized to reproduce and distribute
reprints
for Government purposes notwithstanding any copyright annotations thereon.
The
views and conclusions contained herein are those of the author and should
not be
interpreted as necessarily representing the official policies or
endorsements,
either expressed or implied, of the Defense Advanced Research Projects
Agency,
the Air Force Research Laboratory, or the U.S. Government. (Copyright
2002.)

\section*{Introduction}
In this paper we classify all unitary solutions to the Yang-Baxter 
equation in dimension four. These solutions represent
 both braid operators and quantum operators which describe
 topological entanglement 
and quantum entanglement respectively. The motivating question 
for this work in found in \cite{cqete} in which  Kauffman 
and Lomonaco compare the concepts of entanglement in
the areas of topology and quantum computing. They  
suggest that quantum entanglement may be described in relation to a
 set of entangled
links.

 This classification describes all
$ 4  \times 4 $ unitary matrices that are both
 braid operators and quantum operators.  
Determining  all unitary, $ 4 \times 4 $ matrix solutions to the
Yang-Baxter equation will assist us in studying the relationship
between these two forms of entanglement.  

We introduce  
quantum operators and 
braid operators in Section \ref{oper}.
 We also discuss the connection between the Yang-Baxter 
equation and braids in this section. In Section \ref{ybe}, we
 examine the algebraic
properties of the Yang-Baxter equation. 
We will apply these properties to determine unitary solutions to the
Yang-Baxter equation. 

In Section \ref{matrix}, we describe these properties in terms of matrices. 
 We recall facts about matrices that are
conjugate to unitary matrices. We state the equation used to determine 
the unitary matrices  in Proposition \ref{test} 
and prove a variety of facts chosen to simplify 
the proof of Theorem \ref{uybe}.

In Section \ref{soln}, we present a complete list of unitary $ 4 \times 4 $
matrix solutions to the algebraic Yang-Baxter equation that consists
of five families of unitary matrices. These families are conjugacy
classes determined by matrices that preserve unitarity and 
the property that the conjugate is a solution to the Yang-Baxter 
equation. However, the four families are conjugate under a larger
class of matrices that do not necessarily preserve these properties.
 We obtain the unitary families by analyzing
each solution given in \cite{hier} and applying facts from the earlier
sections. In Section \ref{brac},
 we determine which unitary solutions are also solutions 
to the bracket skein equation given in \cite{knotphys}.

\section{Braid Operators and Quantum Gates} \label{oper}

Topological entanglement and quantum entanglement are non-local structural
 features that occur in topological and quantum systems, respectively.
 Kauffman 
introduces this fact in \cite{cqete} and asks the following question.
What is the relationship between these two concepts? We describe these 
concepts as an introduction to the Yang-Baxter equation. 

Topological entanglement is described in terms of link diagrams and
via the Artin braid group.
The $ \emph{Artin braid group}$ on n-strands is denoted by $ B_n $ and  is
generated by $ \lbrace \sigma_i | 1 \leq i \leq n-1 \rbrace $. The group 
$ B_n $ consists of all words of the form $ \sigma_{j_1} ^{ \pm 1} 
 \sigma_{j_2} ^{ \pm 1} ... \sigma_{j_n} ^{ \pm 1} $ modulo the following 
relationships. 
$$
  \sigma_{i}  \sigma_{i+1} \sigma_{i} = \sigma_{i+1}
 \sigma_{i} \sigma_{i+1}   \text{ for all } 1 \leq i \leq n-1
$$
$$
   \sigma_{i} \sigma_{j} = \sigma_{j} \sigma_{i} \text{ such that }
   | i - j | > 1
$$

\begin{figure}[hbt]  \epsfysize = 1 in
\centerline{\epsffile{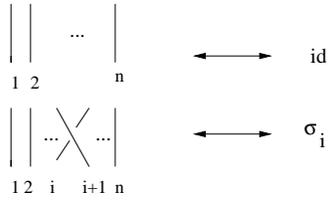}}
\caption{Generators of the Artin Braid Group}
\label{fig:sigma}
\end{figure}

This group has the following diagrammatic representation. A
\emph{n-strand braid} is the immersion of n arcs with over and under crossing 
information at each singularity. Furthermore, any transverse arc intersects the
braid at $ n $ points, except at the levels at which singularities occur.
Each
generator $ \sigma_i $ of the Artin braid group is associated to 
a diagram as shown in Figure \ref{fig:sigma}.
 
\begin{figure} [hbt] \epsfysize = 1 in
\centerline{\epsffile{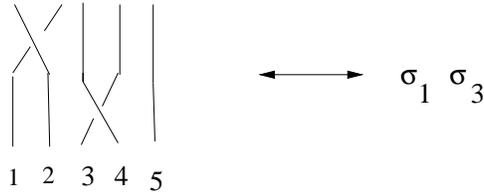}}
\caption{Multiplicaton}
\label{fig:mult}
\end{figure}

Multiplication of diagrams is performed by concatenation of the diagrams from 
upper to lower. We show an example in Figure \ref{fig:mult}.

\begin{figure} \epsfysize = 2 in
\centerline{\epsffile{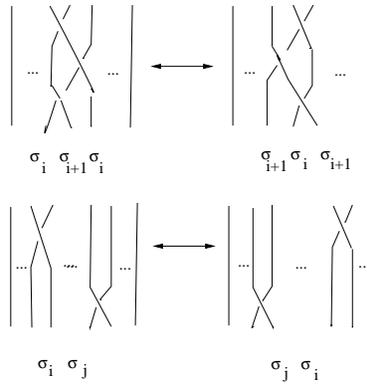}}
\caption{Relations}
\label{fig:rel}
\end{figure}

The diagrammatic form of the relationships in the Artin braid group are 
shown in Figure \ref{fig:rel}.

To obtain the braided Yang-Baxter equation, 
we associate a vector space V to each endpoint in the diagram.
Each n-strand braid represents  a linear map from 
$ V^{\otimes n}  $ to $V^{ \otimes n} $. 
The generator $ \sigma_i $  is associated to a map 
$ R: V \otimes V \rightarrow V \otimes V $ in the following manner:
$ \sigma_i \mapsto  I \otimes I \cdots \otimes R \otimes  \cdots I \otimes I $ with the $ R $ in the ith position and all the $ I $'s representing the identity map. 
This representation will respect the relations of the Artin braid group. 

In the three strand case, this corresponds to the braided Yang-Baxter equation.
\begin{equation} \label{ybe}
 (R \otimes I)(I \otimes R) (R \otimes I) =
 (I \otimes R)(R \otimes I) (I \otimes R)
\end{equation}
Notice that R is a linear map and may be
expressed as a matrix for some basis of V. For the remainder of this
paper we will refer to the Yang-Baxter equation as the ybe.

We algebraically describe quantum entanglement as the action of
of a unitary, linear map with certain properties. 
In a quantum system, quantum states are elements of 
$ V^{ \otimes n} $  where $ V $ is two dimensional vector space over
 $ \pmb{C} $.
 Using Dirac notation, we express the
basis of V as $ \lbrace |0> , |1> \rbrace $. The term 
 $ \emph{qubit}$ refers to  vectors of the form 
$ \alpha |0> + \beta |1> $ such that $ | \alpha |^2 + | \beta |^2 =1 $.
 If $ \alpha $ 
and $ \beta $ are both non-zero then the qubit is said to be in a state of 
$ \emph{superposition} $.  This state collapses
 to  $ |0> $ with probability $ | \alpha |^2 $
and the state $ |1> $ with probability $ | \beta |^2 $. The $ \emph{n-qubit}$ 
is an element of $ V^{ \otimes n} $. If the n-qubit is in superposition, it may simultaneously
 represent $ 2^n $ states of classical information  
with associated probabilities.  The n-qubit will collapse to 
a single outcome when measured, but until then will carry information about
all $ 2^n $  states.

A $ \emph{quantum gate} $ acts on a qubit so that the sum of the 
probabilities of all states is preserved.
 In quantum computing, we assemble a series
of quantum gates to perform a computation. Each quantum gate may be represented
as a unitary matrix.

Consider the case of a two-qubit. This is a quantum state of the 
form: $ \alpha |0> \otimes |0> + \beta  |0> \otimes |1> + \gamma 
 |1> \otimes |0> + \delta  |1> \otimes |1> $. This state is said to 
be $ \emph{entangled} $
 if it can not be written in the form: $ ( x |0> + y |1>)
( \hat{x} |0> + \hat{y} |1> ) $. Note that this implies that $ \alpha \delta 
- \beta \gamma \neq 0 $. We may transform an unentangled state
into a entangled state by application of a quantum gate. In general, 
a n-qubit state $ \psi $ is said to be $ \emph{entangled} $ if 
$ \psi $ can not be written in the form $ P ( \psi_1 \otimes \psi_2) $ where 
$ P: V^{ \otimes n} \rightarrow V^{\otimes n} $ is a permutation of the 
tensor factors of $ V^{\otimes n} $ and $ \psi_1 \in V^{ \otimes l}$, 
$ \psi_2 \in V^{ \otimes k}$ such that $ l + k =n $.

The relationship between topological and quantum entanglement
 is not fully understood.
We examine the properties of the Yang-Baxter equation in 
the next section.

\section{The Yang-Baxter Equation} \label{ybe}
Let V be a vector space over  a field $ F $.
 Let R be a linear map:
$$ 
  R: V \otimes V  \rightarrow V  \otimes V
$$
Let $ I $ be the identity map on $ V $ then $ R $ is said to be a solution
 the \emph{braided Yang-Baxter equation} if the following holds:
$$
  ( R \otimes I) (I \otimes R) (R \otimes I) =
  (I \otimes R) ( R \otimes I) (I \otimes R)
$$
Now assume that V is finite dimensional and that $ \lbrace v_0,v_1, 
\cdot \cdot \cdot  v_{n-1} \rbrace $ is a basis for $ V $ over $ F $.
 We may denote
the basis of $ V \otimes V $ as $ \lbrace v_i \otimes v_j | i,j \in 
\lbrace 0,1,2 ... n-1 \rbrace \rbrace$. Using this basis we may
 describe $ R $ by its action on the generators of $ V \otimes V $:
$$
        R ( v_i \otimes v_j ) = 
        \underset{k,l}{\sum} R^{kl}_{ij} v_k \otimes v_l 
$$
Applying the original definition of the Yang-Baxter equation, we may 
describe $ R $ as a series of equations that are dependent on
the choice of $ i,j,k $ and $x,y,z$.
$$
        \underset{a,b,c}{\sum} R_{ij}^{ab} R_{bk}^{cz} R_{ac}^{xy}
        =
        \underset{m,n,p}{\sum} R_{jk}^{np} R_{in}^{xm} R_{mp}^{yz}
$$
        We prove the following propositions about solutions to the braided 
Yang-Baxter
Equations. These facts will be used to obtain unitary solutions to the
braided 
Yang-Baxter equation. These propositions also indicate that a single solution 
 produces a class of solutions by conjugation and 
scalar multiplication \cite{hier}, \cite{kass}. 

\begin{prop}
 If $ R $ is a solution to the braided 
Yang-Baxter equation then the following hold. \\
 i) If $ \alpha \in F $, then 
 $ \alpha R $ is a solution to the braided Yang-Baxter equation. \\
 ii) If $ Q $ is
 an invertible map, $ Q: V \rightarrow V $,
  then $ (Q \otimes Q) R (Q \otimes Q)^{-1} $
 is a solution to the braided Yang-Baxter equation. \\
 iii) If $ R $  is invertible, 
 then $ R^{-1} $ is a solution to the braided Yang-Baxter equation.
\end{prop}
\noindent
Proof: See \cite{kass} page 168. \qed

We now define the \emph{algebraic Yang-Baxter equation}.
 Let $ \tau : V \otimes V  \rightarrow
         V \otimes V $ be the map such that 
$ \tau (u \otimes v) = (v \otimes u)$. 
Let $ R_{12}, R_{13}, R_{23} : V \otimes V \rightarrow V \otimes V $  
be linear maps. Let $ R_{12} = (R \otimes I)$, 
$ R_{13} = (I \otimes \tau)( R \otimes I) ( I \otimes \tau)$, and
 $ R_{23} = (I \otimes R) $. Then R 
is a solution to the algebraic Yang-Baxter equation if:
\begin{align}\label{aybe}
        R_{12} R_{13} R_{23} = R_{23} R_{13} R_{12}
\end{align}

In this paper, we will refer to the equation as the algebraic  
Yang-Baxter equation and do not follow the notation of \cite{hier}. 
In \cite{hier}, the equation \label{aybe} is refered to as the 
quantum Yang-Baxter equation.
We obtain
a complete list of solutions to the algebraic Yang-Baxter equation from 
\cite{hier}. The algebraic Yang-Baxter equation  is related to  
the braided ybe by the linear transformation  $ \tau $ of $ V \otimes V $.

\begin{prop} If $ R $ is a solution to the braided Yang Baxter equation, 
and $ \tau : V \otimes V \rightarrow V \otimes V $ such that 
$ \tau (u \otimes v) = v \otimes v $ then $ R \circ \tau $ is a
solution to the algebraic Yang-Baxter equation. Similarly, if 
$ \rho $ is a solution to the algebraic Yang-Baxter equation, then 
$ \rho \circ \tau $ is a solution to the braided Yang-Baxter equation.
\end{prop}
Proof:
See \cite{kass}. \qed

Note that if we obtain a solution to the algebraic Yang-Baxter equation,
we may apply $ \tau $ to obtain a solution to the braided
Yang-Baxter equation.

Using the basis for $ V \otimes V $, we rewrite equation \ref{aybe} as a sum
of basis elements. 
\begin{align*}
    R_{12} R_{13} R_{23} (v_i \otimes v_j \otimes v_k)  &=   
         R_{12} R_{13} (I \otimes R) (v_i \otimes v_j \otimes v_k)  \\
        &=  R_{12} R_{13} \underset{b,c}{\sum}R_{j,k}^{b,c}
         (v_i \otimes v_b \otimes v_c)  \\
         &= R_{12} (I \otimes \tau) (R \otimes I)
        \underset{b,c}{\sum}R_{j,k}^{b,c}(v_i \otimes v_c \otimes v_b)  \\
         &= R_{12}(I \otimes \tau) 
        \underset{x,a,b,c}{\sum} R_{j,k}^{b,c} R_{i,c}^{x,a}
        (v_x \otimes v_a \otimes v_b) \\
         &=  R_{12} \underset{x,a,b,c}{\sum} R_{j,k}^{b,c} R_{i,c}^{x,a}
        (v_x \otimes v_b \otimes v_a) \\
         &=  \underset{y,z,x,a,b,c}{\sum} R_{k,j}^{b,c} R_{i,c}^{x,a}
        R_{b,a}^{y,z} (v_x \otimes v_y \otimes v_z)
\end{align*}
Consider the right hand side of the equation and obtain:
\begin{align*}
     R_{23} R_{13} R_{12} (v_i \otimes v_j \otimes v_k) & =
        R_{23} R_{13} \underset{m,n} \sum R_{i,j}^{m,n}
         (v_m \otimes v_n \otimes v_k) \\
        &= R_{23} (I \otimes \tau) (R \otimes I)  \underset{m,n} \sum
        R_{i,j}^{m,n}
         (v_m \otimes v_n \otimes v_k) \\
        &=  \underset{x,y,z,m,n,p}{\sum} R_{i,j}^{m,n} R_{m,k}^{p,x}
        R_{p,n}^{y,z} (v_x \otimes v_y \otimes v_z)
\end{align*}
Therefore, a solution to the algebraic Yang-Baxter equation satisfies the
following equations: 
$$
   \underset{a,b,c}{\sum} R_{k,j}^{b,c} R_{i,c}^{x,a}
        R_{b,a}^{y,z} (v_x \otimes v_y \otimes v_z) =
  \underset{m,n,p}{\sum} R_{i,j}^{m,n} R_{m,k}^{p,x}
        R_{p,n}^{y,z} (v_x \otimes v_y \otimes v_z).
$$

\section{Matrix Representation}\label{matrix}
We may represent solutions to the braided Yang-Baxter equation and the
 algebraic 
Yang-Baxter equation as matrices. We study matrix notation for the 
case where the dimension of $ V $ is two. In this section, we present 
facts about matrix solutions to the Yang-Baxter equation and facts about
unitary matrices. 
  We present a method that determines
  families of unitary solutions to the Yang-Baxter equation.
Recall that we may conjugate $ R $ by $ Q \otimes Q $ or multiply by 
a scalar. The representation of $ Q \otimes Q $ has a specific matrix form 
that will assist in determining the unitary solutions.

Let $ V $ be finite dimensional. If $ R $ is a solution to the quantum 
Yang-Baxter equation, we may rewrite $ R $ as a matrix $ \pmb{R} $. 

Let $  R $ be a solution to the algebraic Yang-Baxter equation, and recall 
that:
$$
        R(v_i \otimes v_j) = 
         \underset{ a,b}{\sum}R_{ij}^{ab} v_a \otimes v_b
$$
$ R $ may be written in matrix form $ \pmb{R}_{ij}^{ab} $ where $ ij $
represents the column and $ ab $ the row of the matrix. In particular,
if the dimension of $ V $ is two then $ V $ 
 has a basis $ \lbrace v_0 , v_1 \rbrace $.
 Hence
$ V \otimes V $ has a basis of the form $ \lbrace
 v_{00}, v_{01}, v_{10}, v_{11} \rbrace $.

$$
\pmb{R} =       \begin{bmatrix}
         R_{00}^{00} &  R_{01}^{00} & R_{10}^{00} & R_{11}^{00}  \\
        R_{00}^{01} &  R_{01}^{01} & R_{10}^{01} & R_{11}^{01} 
         \\ R_{00}^{10} &  R_{01}^{10} & R_{10}^{10} & R_{11}^{10}
          \\ R_{00}^{11} &  R_{01}^{11} & R_{10}^{11} & R_{11}^{11}
  \end{bmatrix}
$$
The matrix \pmb{R} acts on a vector of the form 
$$
\begin{bmatrix}
         a \\  b \\ c \\ c \end{bmatrix}
$$ 
that represents $ a(v_0 \otimes v_0) + b (v_0 \otimes v_1)
 + c (v_1 \otimes v_0) + d (v_1 \otimes v_1) $.

Suppose $ Q: V \rightarrow V $ is an invertible linear map. 
We may rewrite Q as the matrix $ \pmb{Q} $:
$$
        \begin{bmatrix} Q_{0}^{0} & Q_{1}^{0} \\
                        Q_{0}^{1} & Q_{1}^{1} \end{bmatrix}
        \text{ such that } Q_{0}^{0} Q_{1}^{1} - Q_{1}^{0} Q_{0}^{1} \neq 0
$$
Let A denote the linear map $ Q \otimes Q: V 
\otimes V \rightarrow V \otimes V $. This map may be represented as a matrix
$ \pmb{A} $. In the case where the dimension of $ V $ is two, 
 $ \pmb{A} $ is a  $ 4 \times 4 $ matrix
of the form:
$$
        \begin{bmatrix} Q_{0}^{0} Q_{0}^{0} &  Q_{0}^{0}Q_{1}^{0} &
          Q_{1}^{0}Q_{0}^{0} & Q_{1}^{0} Q_{1}^{0} \\
                        Q_{0}^{0} Q_{0}^{1} & Q_{0}^{0} Q_{1}^{1} &
                  Q_{1}^{0} Q_{0}^{1} & Q_{1}^{0}  Q_{1}^{1} \\
                 Q_{1}^{0} Q_{0}^{0} & Q_{1}^{0} Q_{1}^{0} &
         Q_{1}^{1} Q_{0}^{0} &  Q_{1}^{1} Q_{1}^{0} \\
                        Q_{1}^{0} Q_{0}^{1} &  Q_{1}^{0} Q_{1}^{1} &
                 Q_{1}^{1} Q_{0}^{1} & Q_{1}^{1} Q_{1}^{1}      
        \end{bmatrix}
$$

For each matrix solution in \cite{hier}, there is a family of solutions 
to the algebraic Yang-Baxter equations. To obtain another element of this
family, we conjugate $ \pmb{R} $ by $ \pmb{A} $. We wish to obtain all
unitary families of solutions. Note that it is not 
sufficient to determine which representatives are unitary. 
 Although $ \pmb{R} $ may not be unitary, it is possible that 
$ \pmb{ARA}^{-1} $ is unitary for some $ \pmb{A}$.

\section*{Linear Algebra Facts}
We prove facts about matrix solutions to 
the algebraic Yang-Baxter equation. These facts determine a preliminary 
approach to evaluating whether or not a family of solutions will produce
any unitary solutions. 
Let $ M $ be a finite dimensional matrix. $ M^{T} $ denotes the transpose 
of the matrix $ M $. Let $ \bar{M} $ denote the conjugate of $ M $. 
Let $ M ^{*} $ denote the conjugate transpose of the
matrix $ M $. Recall that $ M $ is a \emph{unitary} matrix if
 $ M^{*} = M^{-1} $.
\begin{prop}
Recall the following matrix facts. \\
i)Let A be an invertible matrix then  $ (A^{*})^{-1} = (A^{-1})^* $. \\
ii)If A is a unitary matrix and let $ \alpha \in \Bbb{C} $,
then $ \alpha $A is a unitary matrix if and only if $ | \alpha | = 1 $. \\ 
iii) If A is a unitary matrix and $ \lambda $ is an eigenvalue 
of A, then $ | \lambda |  = 1 $. \\
\end{prop}

\begin{prop} \label{test} If $ ARA^{-1} $ is unitary, then \\
 i)  $ AR^{-1}A^{-1} = 
(A^*)^{-1} R^* A^* $ implying that
  $ A^* A R^{-1} = R^* A^* A $, and \\
 ii)  $ R^{-1} $ and $ R^* $
  have the same set of eigenvalues. 
\end{prop}
\noindent
Proof: Suppose $ ARA^{-1} $ is unitary. By the definition of unitary:
\begin{align*}
        (ARA^{-1})^{-1} &= (ARA^{-1})^{*} \\
        A R^{-1} A^{-1}  & = (A^{-1})^{*} R^{*} A^{*} \\
        A R^{-1} A^{-1} & =  (A^{*})^{-1} R^{*} A^{*} \\
        A^{*}A R^{-1} & = R^{*} A^{*} A
\end{align*}

$ R^{-1} $ and $ R^{*}$ are conjugate by $ A^{*} A $. Therefore, they
have the same set of eigenvalues. \qed

We make the following observations about an invertible 
 matrix $ Q $ of the form
$$
 Q= \begin{bmatrix} a & b \\
                    c & d \end{bmatrix}
$$
\begin{prop}\label{diag}If 
$$ c = - \frac{a \bar{b}}{\bar{d}} 
$$ then $ Q^* Q $ is diagonal.
If $ A= Q \otimes Q $ then $ A^* A $ is also diagonal. 
\end{prop}
\noindent
Proof: Recall that the matrix form of $ Q \otimes Q $ is:
$$ A =
\begin{bmatrix} aa & ab & ba & bb \\ 
        ac & ad & bc & bd \\
        ca & cb & da & db \\  
        cc & cd & dc & dd 
\end{bmatrix}
$$
Let $ H = A^* A $.  Let $ x = (a \bar{a} + c \bar{c}) $, 
$ y = (b \bar{b} + d \bar{d}) $, and $ z = (a \bar{b} + c \bar{d})$.
$$
H = \begin{bmatrix} x^2 & x \bar{z} & \bar{z}x &  \bar{z}^2 \\
                    xz & xy & \bar{z}z & \bar{z} y \\
                    zx & z \bar{z} & yx & y \bar{z} \\
                    z^2 & zy & yz & y^2 \\      
\end{bmatrix}
$$ 
If
$$  c = - \frac{a \bar{b}}{\bar{d}} $$
then $ z = 0 $ and $ H $ is a diagonal matrix. \qed

The relationship between a matrix $ Q $ of this form
and a $ 4 \times 4 $ matrix $ B $ may be simplified using the 
following facts. 
\begin{prop}\label{test} Let $ H = A^* A $ and  let $ D = H B^{-1} - B^* H$.
Then
$$ 
        D_{ij} = \underset{k}{\sum} H_{ik}B^{-1}_{kj} - \underset{m}{ \sum} B^*_{im} H_{mj}
$$
If $ H $ is a diagonal matrix, then
$$ 
        D_{ij} =  H_{ii}B^{-1}_{ij} -  B^*_{ij} H_{jj}
$$
\end{prop}

We will apply these facts to determine unitary solutions to the
algebraic Yang Baxter equation. 

\begin{prop}\label{contra} If $ H $ is constructed as 
before then $ H_{ii} \neq 0 $ for all $ i $. 
\end{prop}
\noindent
Proof: If $ H_{i,i} = 0 $ for some $ i $, then either $ x =0 $ or $ y =0 $. 
This contradicts
 the assumption that $ Q $ is an invertible matrix.\qed

\begin{prop}\label{hermite} Suppose H is constructed as before. 
If $ H_{ij} = 0 $ for some $ i \neq j$ then 
 $ H_{ij} = 0 $ for all $ i\neq j $ 
\end{prop}
\noindent
Proof:   We observe that 
$$
 H_{ij} = \underset{k}{\sum} A^*_{ik} A_{kj} . $$ 
      
 If $ H_{ij} = 0 $ for some $ i  \neq j $ then 
 we observe  that $ x = 0 $,$ y = 0 $, or
 $ z = 0 $. If $ x = 0 $ then $ |a|^2 + |c|^2 = 0 $ implying that 
$ a = c = 0 $ and $ Q $ is not invertible. This 
 contradicts our definition of $ Q $.  We have a similar contradiction if 
$ y =0 $. Note that $ z $ is a factor of each off diagonal entry. We have
 eliminated the the other possiblities.
 Therefore $ x \neq 0, y \neq 0$ and $  z = 0 $ which forces 
H to be a diagonal matrix. \qed

We apply these facts to the families of solutions in \cite{hier}. If 
the matrix $ \pmb{R} $
produces unitary solutions under conjugation, then 
$ \pmb{R} $ is of rank 4 and $ \pmb{R}^* $ and 
$ \pmb{R}^{-1} $ have equivalent sets of eigenvalues.
Further, if $ \pmb{ARA}^{-1} $ is a solution to the Yang-Baxter equation, 
then $ \pmb{A}^* \pmb{AR} = \pmb{R}^{-1} \pmb{A}^*  \pmb{A} $.
If $ \pmb{R} $ is a unitary solution to the quantum Yang-Baxter, then
$ \alpha \pmb{R} $ is a unitary solution if and only if $ | \alpha | =1 $.

\section{Solutions to the Yang-Baxter Equation}\label{soln}
\begin{thm}\label{uybe}

There are five families of 4 $\times $ 4 unitary matrix 
solutions to the braided Yang-Baxter equation.
Each solution has the form:
$$
        \it{k}  A R  A^{-1} T 
$$
where \it{k} is a scalar of norm one, $ Q $ is an invertible matrix 
such that: 
\begin{align*} 
Q &= \begin{bmatrix} a & b \\
                     c & d \end{bmatrix}  \\
A = Q \otimes Q &= \begin{bmatrix} a^2 & ab & ba & b^2 \\
                     ac & ad & bc & bd \\
                     ca & cb & da & db \\
                     c^2 & cd & cd & d^2 \end{bmatrix} \\
\end{align*}
and 
\begin{align*}
T &= \begin{bmatrix} 1 & 0 & 0 & 0\\
                          0 & 0 & 1 & 0\\
                          0 & 1 & 0 & 0\\
                          0 & 0 & 0 & 1\\
                  \end{bmatrix} \\
\end{align*}
Additional information about the matrices $ R $ and $ A $ are
 specified for each family. \\
\noindent
Family 1: \\

The matrix $ R $ has the form
\begin{align} \label{fam1}
R &= \begin{bmatrix} 1 & 0 & 0 & 0\\
                          0 & p & 0 & 0\\
                          0 & 0 & q & 0\\
                          0 & 0 & 0 & r\\
                  \end{bmatrix}  \text { and}
\end{align}
\begin{align*}
  1 &= p \bar{p} = q \bar{q} = r \bar{r}. 
\end{align*}
The variable $ c $ in the matrix Q has the restriction that
\begin{align*}
  c &= - {\displaystyle \frac {a\,\overline{b}}{\overline{d}}}. 
\end{align*} \\
 
\noindent
Family 2: \\

The matrix R has the form:
\begin{align}\label{fam2}
R &= \begin{bmatrix} 0 & 0 & 0 & p\\
                          0 & 0 & 1 & 0\\
                          0 & 1 & 0 & 0\\
                          q & 0 & 0 & 0\\
                  \end{bmatrix}  \text{ and}
\end{align}
when $ c \neq - \frac{a \bar{b}}{ \bar{d}} $ and
\begin{align*}
   |pq| &= 1,   \\
  p &=\frac{(b \bar{b} + d \bar{d})(\bar{a} b + \bar{c} d)}{
   (a \bar{a} + c \bar{c })(a \bar{b} +c \bar{d})}, \\
  q &=\frac{(a \bar{a} + c\bar{c})(a \bar{b} +c \bar{d})
    }{(b \bar{b} + d\bar{d})(\bar{a} b + \bar{c} d) }.
\end{align*} \\
\noindent
Family 3: \\

The third family consists of matrices with the form:
\begin{align}\label{fam3}
R &= \begin{bmatrix} 0 & 0 & 0 & p\\
                          0 & 0 & 1 & 0\\
                          0 & 1 & 0 & 0\\
                          q & 0 & 0 & 0\\
                  \end{bmatrix} \text{ and}
\end{align}
\begin{align*}
   |pq| &= 1,   \\ 
  p \bar{p} &= \frac{ (d \bar{d})^2}{ (a \bar{a})^2}, \\
  q \bar{q} &= \frac{(a \bar{a})^2 }{ ( d \bar{d})^2}.
\end{align*}
The matrix $ Q $ has the restriction that
\begin{align*}
 c &= \frac{- a \bar{b} }{ \bar{d}}. \\
\end{align*} \\
\noindent
Family 4: \\
The matrix $ R $ has the form
\begin{align}\label{fam4}
R &= \begin{bmatrix} \frac{1}{ \sqrt(2)} & 0 & 0 & \frac{1}{ \sqrt(2)} \\
                          0 & \frac{1}{ \sqrt(2)}  & \frac{1}{ \sqrt(2)}  & 0\\
                          0 & \frac{1}{ \sqrt(2)}  & -\frac{1}{ \sqrt(2)}  & 0\\
                          -\frac{1}{ \sqrt(2)}  & 0 & 0 & \frac{1}{ \sqrt(2)} \\
                  \end{bmatrix}. 
\end{align}
The matrix $ Q $ has the following restrictions:
\begin{align*}
  c &= - \frac{a \bar{b}}{ \bar{d}}\\
   |a| &= |d|.
\end{align*}

\noindent
Family 5: \\
The matrix $ R $ has the form
\begin{align}\label{fam5}
R &= \begin{bmatrix} 1 & 0 & 0 & 0\\
                          0 & 0 & 1 & 0\\
                          0 & 1 & 0 & 0\\
                          0 & 0 & 0 & 1\\
                  \end{bmatrix}. 
\end{align}
There are no additional restrictions on the matrix $ Q $.
\end{thm}  
\begin{rem}
We may transform the elements of families 2 and 3 into elements from 
family 1 by conjugation by unitary matrices. However, these matrices do not 
have the form $ Q \otimes Q $. 
\end{rem}
\begin{rem}
The matrices of families 2 and 3 perform quantum entanglement. The matrices from families 2 and 3 do not detect knotting but do detect linking. Family 4 performs quantum entanglement  and detects knotting and linking \cite{ls}.
\end{rem}
\noindent
Proof:
If $ \phi $ is a solution to 
the algebraic Yang-Baxter equation and $ \tau: V \otimes V
 \rightarrow V \otimes V $ is a 
linear transformation such that $ \tau(v_i \otimes v_j) = v_j \otimes v_i $,
then $ \phi \circ \tau $ is a solution to the braided Yang Baxter equation.
We may express $ \tau $ as the unitary matrix $ T $
since $ V \otimes V $ has a basis $ \lbrace v_0 \otimes v_0 ,
v_0 \otimes v_1,  v_1 \otimes v_0 , v_1 \otimes v_1  \rbrace$.

Note that $ R $ is a unitary solution to the algebraic Yang-Baxter equation 
if and only if 
$ R T $ is a unitary solution to the braided Yang-Baxter equation. 

To find all unitary solutions to the braided Yang-Baxter equation, it suffices
to determine 
all unitary solutions to the algebraic Yang-Baxter and multiply by $ T $. 

The families of solutions to the algebraic Yang-Baxter equation  in \cite{hier}
 are indicated by a single element. This representative element is
is conjugated by the matrix $ Q \otimes Q $
 and multipled by a scalar $ \it{ k } $ to 
produce the entire family.
If a family of solutions produces any unitary solutions, the representative
of the family must be invertible  
From \cite{hier} we obtain the 
following list of invertible representatives: 
\begin{align*}
 \mathit{R01}&= \begin{bmatrix}
1 & 0 & 0 & 1 \\
0 & -1 & 0 & 0 \\
0 & 0 & -1 & 0 \\
0 & 0 & 0 & 1
\end{bmatrix} \\ 
\mathit{R02}&= \begin{bmatrix}
1 & 0 & 0 & 1 \\
0 & 1 & 1 & 0 \\
0 & 1 & -1 & 0 \\
-1 & 0 & 0 & 1
\end{bmatrix}  \\
\mathit{R03} &= \begin{bmatrix}
1 & 0 & 0 & 0 \\
0 & 0 & 1 & 0 \\
0 & 1 & 0 & 0 \\
0 & 0 & 0 & 1
\end{bmatrix} 
\end{align*}
\begin{align*}
\mathit{R11} &= \begin{bmatrix}
p^{2} + 2\,p\,q - q^{2} & 0 & 0 & p^{2} - q^{2} \\
0 & p^{2} + q^{2} & p^{2} - q^{2} & 0 \\
0 & p^{2} - q^{2} & p^{2} + q^{2} & 0 \\
p^{2} - q^{2} & 0 & 0 & p^{2} - 2\,p\,q - q^{2}
\end{bmatrix} \\
\mathit{R12} &= \begin{bmatrix}
p & 0 & 0 & k \\
0 & p & p - q & 0 \\
0 & 0 & q & 0 \\
0 & 0 & 0 &  - q
\end{bmatrix}  \\
\mathit{R13} &= \begin{bmatrix}
k^{2} & k\,p &  - k\,p & p\,q \\
0 & k^{2} & 0 & k\,q \\
0 & 0 & k^{2} &  - k\,q \\
0 & 0 & 0 & k^{2}
\end{bmatrix}\\
\mathit{R14} &= \begin{bmatrix}
0 & 0 & 0 & p \\
0 & 0 & k & 0 \\
0 & k & 0 & 0 \\
q & 0 & 0 & 0
\end{bmatrix}
\end{align*}
\begin{align*} 
\mathit{R21} &= \begin{bmatrix}
k^{2} & 0 & 0 & 0 \\
0 & k\,p & k^{2} - p\,q & 0 \\
0 & 0 & k\,q & 0 \\
0 & 0 & 0 & k^{2}
\end{bmatrix}\\
\mathit{R22} &= \begin{bmatrix}
k^{2} & 0 & 0 & 0 \\
0 & k\,p & k^{2} - p\,q & 0 \\
0 & 0 & k\,q & 0 \\
0 & 0 & 0 &  - p\,q
\end{bmatrix}\\
\mathit{R23} &= \begin{bmatrix}
k & p & q & s \\
0 & k & 0 & q \\
0 & 0 & k & p \\
0 & 0 & 0 & k
\end{bmatrix}  \\
\mathit{R31} &= \begin{bmatrix}
k & 0 & 0 & 0 \\
0 & p & 0 & 0 \\
0 & 0 & q & 0 \\
0 & 0 & 0 & s
\end{bmatrix}
\end{align*}
If $ R $ produces unitary solutions, then $ R^* $ and $ R^{-1} $ are
conjugate. 
We eliminate from this list all matrices that do not satisfy the condition
that  $ R^* $ and $ R^{-1} $ have the same set of eigenvalues.
From this computation, we determine only the representatives:
$ R01, R02, R03, R12, R13, R14, R21, R22, R23, $ and $ R31 $
 could produce unitary solutions
to the Yang-Baxter equation. The eigenvalues in each of these matrices
 are non-zero, using this fact we have removed one degree of freedom
from each representative. Multiplying by a non-zero eigenvalue determines
the new representatives used in the following cases. 
To determine if a representative produces unitary solutions, 
we refer to Proposition \ref{test}. If $ R $ and $ Q $ produce a unitary 
matrix, then $ D_{ij}=0 $ for all $i,j$.

\section*{Case: R31}  
We consider the case of the 
diagonal matrix $ R31 $. 
Note that if $ R31^* $ and $ R31^{-1} $
have the same set of eigenvalues then 
$ R31 $ is unitary and each eigenvalue
has norm one. Further, the eigenvalues are the diagonal elements. 
$$
R31 = \begin{bmatrix} 1 & 0 & 0 & 0 \\
                 0 & p & 0 & 0 \\
                 0 & 0 & q & 0 \\
                0 & 0 & 0 & k 
\end{bmatrix}.
$$
The inverse and conjugate transpose of the matrix are: 
$$
R31^* = \begin{bmatrix} 1 & 0 & 0 & 0 \\
                 0 & \bar{p} & 0 & 0 \\
                 0 & 0 & \bar{q} & 0 \\
                0 & 0 & 0 & \bar{k} \end{bmatrix}
R31^{-1} = \begin{bmatrix} 1 & 0 & 0 & 0 \\
                 0 & \frac{1}{p} & 0 & 0 \\
                 0 & 0 & \frac{1}{q} & 0 \\
                0 & 0 & 0 & \frac{1}{k}  \end{bmatrix}.
$$

To determine the family of solutions given by $ R31 $
we examine the equation from Proposition \ref{test}.
$$
   D_{ij} = \sum{k} H_{ik}R31^{-1}_{k,j} - \underset{m}{ \sum} R31^*_{im} H_{mj}
$$
To obtain a unitary R matrix under conjugation by $ Q $, each
Each $ D_{ij} = 0 $.
$ R31 $ is a diagonal matrix and by Proposition \ref{diag}
\begin{align*}
  D_{ij} &=  H_{ij}R31^{-1}_{jj} -  R31^*_{ii} H_{ij} \\
      0  &= (R31^{-1}_{jj} - R31^{-1}_{ii}) H_{ij} \\
\end{align*}
From this equation, we obtain two cases. 
In case one,  $ R31^{-1}_{jj} - R31^{-1}_{ii} = 0 $ for all $i,j $.
This implies that $ R31 $ is the identity matrix, and we obtain
solutions that are scalar multiples of the identity matrix.

In the second case,  $ R31 $ is not a multiple of the identity.
If $ R31 $ is not a multiple of the identity, then 
$$
(R31^{-1}_{jj} - R31^{-1}_{ii}) \neq 0 \text{ for some } i \neq j.
$$
Therefore, 
$ H_{ij} = 0 $ for $ i \neq j $. By Proposition \ref{hermite},
$ H $ is a diagonal matrix and  
$$
c = \frac{-a \bar{b} }{ \bar{d} }
$$
This produces the first family of solutions.

\section*{Case: $ R21 $ }

The matrix $ R21 $ is not unitary. We observe that $ R21^* $ 
and $ R21^{-1} $ have the same set of eigenvalues if $ |p| = |q|= 1 $. 

$$
R21 = \begin{bmatrix} 1 & 0 & 0 & 0 \\
                 0 & p & 1-pq & 0 \\
                 0 & 0 & q & 0 \\
                0 & 0 & 0 & 1 
\end{bmatrix}
$$
The inverse and conjugate transpose of the matrix R21 are: 
$$
R21^* = \begin{bmatrix} 1 & 0 & 0 & 0 \\
                 0 & \bar{p} & 0 & 0 \\
                 0 & 1-\bar{pq} & \bar{q} & 0 \\
                0 & 0 & 0 & 1 \end{bmatrix}
R21^{-1} = \begin{bmatrix} 1 & 0 & 0 & 0 \\
                 0 &\frac{1}{p} & \frac{-1}{pq} +1 & 0 \\
                 0 & 0 & \frac{1}{q} & 0 \\
                0 & 0 & 0 & 1 \end{bmatrix}
$$

To determine when $ R21 $  produces unitary solutions under conjugation, 
we examine: 
$$
  D_{ij} = \sum{k} H_{ik} R21^{-1}_{kj} - \underset{m}{ \sum } R21^*_{im} 
H_{mj}
$$

If $ H $ and $ R21 $ produce a unitary solution, then $ D $ is the zero matrix.
Examine the entry 
\begin{align*}
  D_{12} &= \underset{k}{ \sum } H_{1k}R21^{-1}_{k2} - \underset{m}{\sum }
         R21^*_{1m} H_{m2} \\
          0 &= H_{12} \frac{1}{p} - H_{12} \\
          0 &= (\frac{1}{p}-1) H_{12} \\   
\end{align*}
From this equation, we obtain two cases: 
 either $ H $ is a diagonal matrix by Proposition \ref{hermite} or $ p=1 $. 
If $ p=1 $
\begin{align*}
 D_{23} &= \underset{k}{ \sum} H_{2k}R21^{-1}_{k3} 
 - \underset{m}{ \sum} R21^*_{2m} H_{m3} \\
        0 &=  H_{22} (1-\frac{1}{q}) + H_{23}\bar{q}  - H_{23}\bar{q} \\
        0 &= H_{22} (1-\frac{1}{q}) \\
\end{align*}
We conclude that $ q=1 $ since if $ H_{22} = 0 $ then $ Q $ is not
 invertible.
Therefore if $ p=1 $ then $q=1$ which is an element of family 1. 

If H is a diagonal matrix then 
\begin{align*}
  c &= - \frac{a \bar{b} }{ \bar{d}}
\end{align*} implying that 
\begin{align*}
   D_{ij} &= H_{ii}R21^{-1}_{ij} - R21^*_{ij} H_{jj} \\
   D_{23} &= H_{22}R21^{-1}_{23} - R21^*_{23} H_{33} \\
   D_{23} &= H_{22} (- \frac{1}{pq} +1 ) \\
    D_{23} &= H_{22}( -\frac{1}{pq} +1 )
\end{align*}
Hence $  -\frac{1}{pq} +1 =0 $ or $ p = \frac{1}{q}$, which
is a subcase of family 1.
All unitary solutions produced by $ R21 $ are in family 1.

\section*{Case: $ R22 $}
The matrix $ R22 $ is of the following form after scaling. 
Note that $ |p|=|q|=1 $.
 
$$
R22 = \begin{bmatrix} 1 & 0 & 0 & 0 \\
                 0 & p & 1-pq & 0 \\
                 0 & 0 & q & 0 \\
                0 & 0 & 0 & -pq
\end{bmatrix}.
$$
The inverse and conjugate transpose of $ R22 $ have the form
$$
 R22^{-1} = \begin{bmatrix} 1 & 0 & 0 & 0 \\
                 0 & \frac{1}{p} & -\frac{1}{pq}+1 & 0 \\
                 0 & 0 & \frac{1}{q} & 0 \\
                0 & 0 & 0 & -\frac{1}{pq}
\end{bmatrix} \text{ and }
R22^* = \begin{bmatrix} 1 & 0 & 0 & 0 \\
                 0 & \bar{p} & 0 & 0 \\
                 0 & 1-\bar{pq} & \bar{q} & 0 \\
                0 & 0 & 0 & -\bar{pq}
\end{bmatrix}.
$$
To determine if $ R22 $  produces unitary solutions under conjugation, 
we examine: 
\begin{align*}
  D_{ij} & = \underset{k}{ \sum} H_{ik}R22^{-1}_{kj} - \underset{m}{ \sum}
                R22^*_{im} H_{mj} \\
 D_{12} & = \underset{k}{ \sum} H_{1k}R22^{-1}_{k2} - \underset{m}{ \sum}
             R22^*_{1m} H_{m2} \\
             0 &= \frac{1}{p} H_{12} - H_{12} \\
             0 &= (-\frac{1}{p}+1) H_{12} \\ 
\end{align*}
This implies H is a diagonal matrix, by Proposition \ref{hermite}, or $ p=1 $.
Let $ p=1 $, and examine
\begin{align*}
 D_{23} &= \underset{k}{ \sum} H_{2k}R22^{-1}_{k3} - \underset{m}{ \sum}
 R22^*_{2m} H_{m3} \\
        0 &=  H_{22} (1-\frac{1}{q}) + H_{23}\bar{q}  - H_{23}\bar{q} \\
        0 &= H_{22} (1-\frac{1}{q}) \\
\end{align*}
If $ H_{22}=0 $ then $ Q $ is not invertible. 
The solution which consists of 
 $ p = q =1 $ is a subcase of family 1.

If $ H $ is a diagonal matrix, then 
\begin{align*} 
  c &= - \frac{a \bar{b} }{ \bar{d} } \text{ and} \\
  D_{ij} &= H_{ii} R22^{-1}_{ij} - R22^*_{ij} H_{jj} \\
  D_{23} &= H_{22} (-\frac{1}{pq} +1) \\
  0 &= H_{22} (-\frac{1}{pq} +1)
\end{align*}
Hence $ p = \frac{1}{q} $ which is a subcase of family 1.

\section*{The matrix $ R23 $ }
We examine the matrix $ R23 $. We may multiply by a scalar and assume that
$$
\mathit{R23} = \begin{bmatrix}
1 & p & q & s \\
0 & 1 & 0 & q \\
0 & 0 & 1 & p \\
0 & 0 & 0 & 1
\end{bmatrix}
$$
In this case: 
$$
\mathit{R23}^{-1} = \begin{bmatrix}
1 & -p & -q & -s + 2pq \\
0 & 1 & 0 & -q \\
0 & 0 & 1 & -p \\
0 & 0 & 0 & 1
\end{bmatrix} \text{ and }
\mathit{R23}^* = \begin{bmatrix}
1 & 0 & 0 & 0 \\
\bar{p} & 1 & 0 & q \\
\bar{q} & 0 & 1 & p \\
\bar{s} & \bar{q} & \bar{p} & 1
\end{bmatrix}
$$

We obtain 
\begin{align*}
 D_{12}  &= \underset{k}{ \sum} H_{1k}R23^{-1}_{k2} - \underset{m}{ \sum} R23^*_{1m} H_{m2} \\
          0 &= H_{11} (-p) + H_{12} - H_{12} \\
          0 &= -p H_{1,1} \\      
\end{align*}

Note that $ p=0$. If $ H_{11} = 0 $, the matrix $ Q $ is not invertible.
We consider the following two entries.
\begin{align*}
D_{13}  &= \underset{k}{ \sum} H_{1k}R23^{-1}_{k3} - \underset{m}{ \sum} R23^*_{1m} H_{m3} \\
          0 &= - H_{11}q  + H_{13} - H_{13}, \\
          0 &= -q H_{11}. \\      
D_{14}  &= \underset{k}{ \sum} H_{1k}R23^{-1}_{k4} - \underset{m}{ \sum} R23^*_{1m} H_{m4} \\
          0 &= H_{11} (-2 + 2pq) - H_{12}p + - H_{13}q +  H_{14} - H_{14}, \\
          0 &= (-s+2pq) H_{11} - H_{12}p + - H_{13}q. \\        
\end{align*}
Hence $p = q= s = 0 $. R23 produces trivial solutions which are contained
in family 1. 

\section*{Case: $ R12 $ }
After scaling, the matrix $ R12 $ has the form:
$$
\mathit{R12} := \begin{bmatrix}
1 & 0 & 0 & k \\
0 & 1 & 1 - q & 0 \\
0 & 0 & q & 0 \\
0 & 0 & 0 &  - q
\end{bmatrix}
$$
The inverse and the conjugate transpose of the matrix $ R12 $ are:
$$
\mathit{R12}^{-1} = \begin{bmatrix}
1 & 0 & 0 & -\frac{k}{q} \\
0 & 1 & 1 - \frac{1}{q} & 0 \\
0 & 0 & \frac{1}{q} & 0 \\
0 & 0 & 0 &  - \frac{1}{q}
\end{bmatrix} \text{ and }
\mathit{R12}^* = \begin{bmatrix}
1 & 0 & 0 & 0 \\
0 & 1 & 0 & 0 \\
0 & 1-\bar{q} & \bar{q} & 0 \\
\bar{k} & 0 & 0 &  - \bar{q}
\end{bmatrix}.
$$
Observe that 
\begin{align*}
 D_{23}  &= \underset{k}{ \sum} H_{2k}R12^{-1}_{k3} - \underset{m}{ \sum} R12^*_{2m} H_{m3} \\
          0 &= H_{22} (1-\frac{1}{q}) + H_{23} \frac{1}{q}  - H_{23} \\
         0  &= (1-\frac{1}{q}) (H_{22} - H_{23}).    
\end{align*}
Therefore, either $q = 1 $ or $ H_{22} = H_{23}$ .
If $ q=1 $ then
\begin{align*}
 D_{14}  &= \underset{k}{ \sum} H_{1k}R12^{-1}_{k4} - \underset{m}{ \sum} R12^*_{1m} H_{m4} \\
          0 &=  -k H_{11}   + H_{14}   - H_{14} \\
          0 &=  -k H_{11}, \\    
\end{align*}

and $ k =0 $  since $ H_{11} \neq 0 $. This solution is part of family 1.

Suppose that $ H_{22}=H_{23} $ then $ det(Q) = 0 $ 
 contradicting the assumption that $ Q $ was invertible.

\section*{Case: $ R13 $}
The matrix $ R13 $ has the form 
$$
\mathit{R13} :=  \begin{bmatrix}
1 & p & -p & pq \\
0 & 1 & 0 & q \\
0 & 0 & 1 &  - q \\
0 & 0 & 0 & 1
\end{bmatrix}.
$$
The inverse and conjugate transpose of $ R13 $ are
$$
\mathit{R13}^{-1}=  \begin{bmatrix}
1 & -p & p & pq \\
0 & 1 & 0 & -q \\
0 & 0 & 1 & q \\
0 & 0 & 0 & 1
\end{bmatrix} \text{, }
\mathit{R13}^* =  \begin{bmatrix}
1 & 0 & 0 & 0 \\
\bar{p} & 1 & 0 & 0 \\
\bar{-p} & 0 & 1 &  0 \\
\bar{-pq} & \bar{q} & \bar{-q} & 1
\end{bmatrix}.
$$

We consider the equation: 
\begin{align*}
D_{12}  &= \underset{k}{ \sum} H_{1k} R13^{-1}_{k2} - 
        \underset{m}{ \sum} R13^*_{1,m} H_{m2} \\
        0 &= -p H_{11} + H_{12} - H_{12} \\
        0 &= -p H_{11} \\
\end{align*}
$H_{11} \neq 0 $ by Proposition \ref{contra}
If $ p = 0 $ then
\begin{align*}
D_{24}  &= \underset{k}{ \sum} H_{2k} R13^{-1}_{k4} - 
        \underset{m}{ \sum} R13^*_{2m} H_{m4} \\
        0 &= -q H_{22} + q H_{23} + H_{24} - H_{24} \\
        0 &=  -q H_{22} + q H_{23} \\
\end{align*}

If $ H_{22} = H_{23} $ then $ det(Q) = 0 $. Hence, $ q =0 $
and this solution is a subcase of family 1.

\section*{Case: $ R14 $ }
The matrix has the following form after scaling: 
$$
\mathit{R14} =  \begin{bmatrix}
0 & 0 & 0 & p \\
0 & 0 & 1 & 0 \\
0 & 1 & 0 & 0 \\
q & 0 & 0 & 0
\end{bmatrix}
$$
We assume that $ |pq| = 1 $ and that the inverse and conjugate transpose are
$$
\mathit{R14}^{-1} =  \begin{bmatrix}
0 & 0 & 0 &\frac{1}{q} \\
0 & 0 & 1 & 0 \\
0 & 1 & 0 & 0 \\
\frac{1}{p} & 0 & 0 & 0
\end{bmatrix} \text{ and }
\mathit{R14}^* =  \begin{bmatrix}
0 & 0 & 0 & \bar{q} \\
0 & 0 & 1 & 0 \\
0 & 1 & 0 & 0 \\
\bar{p} & 0 & 0 & 0
\end{bmatrix}.
$$
Consider the individual entry
\begin{align*}
D_{43}  &= \underset{k}{ \sum} H_{4k} R14^{-1}_{k3} - 
        \underset{m}{ \sum} R14^*_{4m} H_{m3} \\
        0 &=  H_{42} - \bar{p} H_{13} \\
        0 &=  (b \bar{b} + d \bar{d}) ( a \bar{b} + c \bar{d}) 
            - \bar{p} (a \bar{a} + c \bar{c}) ( \bar{a}b +  \bar{c}d).   
\end{align*}
As a result
$$
 p = 
\frac{ (b \bar{b} + d \bar{d})( \bar{a}b +  \bar{c}d)}
{(a \bar{a} + c \bar{c})( a \bar{b} + c \bar{d}) } 
$$
or 
$$
 c= -\frac{ a \bar{b} }{ \bar{d} }. 
$$

Suppose that $c \neq - \frac{a \bar{b}}{ \bar{d}} $ and $ p = 
\frac{(a \bar{a} + c \bar{c})( \bar{a}b +  \bar{c}d)}
{(b \bar{b} + d \bar{d})( a \bar{b} + c \bar{d})} $
then

\begin{align*}
D_{12}  &= \underset{k}{ \sum} H_{1k} R14^{-1}_{k2} - 
        \underset{m}{ \sum} R14^*_{1m} H_{m2} \\ 
        0 &= H_{13}  - \bar{q} H_{42} \\
        0 &= (a \bar{a} + c \bar{c})( \bar{a}b +  \bar{c}d)
        - \bar{q}(b \bar{b} + d \bar{d})( a \bar{b} + c \bar{d}) \\
\end{align*}
Hence 
$$ 
q = \frac{(a \bar{a} + c \bar{c})( a \bar{b} + c \bar{d}) }
     { (b \bar{b} + d \bar{d})( \bar{a}b +  \bar{c}d)}
$$
Each entry is now zero in the matrix D.
If 
\begin{align*}
p &= 
\frac{ (b \bar{b} + d \bar{d})( \bar{a}b +  \bar{c}d)}
{(a \bar{a} + c \bar{c})( a \bar{b} + c \bar{d}) }  \\
q &= \frac{(a \bar{a} + c \bar{c})( a \bar{b} + c \bar{d}) }
     { (b \bar{b} + d \bar{d})( \bar{a}b +  \bar{c}d)} \\
\end{align*}
we obtain the second family of solutions.

Now suppose that  $ c = - \frac{a \bar{b} }{ \bar{d} } $. 
All entries in D except the following two entries are zero.
\begin{align*}
D_{14}  &= \underset{k}{ \sum} H_{1k} R14^{-1}_{k4} - 
        \underset{m}{ \sum} R14^*_{1m} H_{m4} \\ 
        0 &= H_{11} \frac{1}{q} - \bar{q} H_{44} \\
        0 &= (a \bar{a} + c \bar{c})^2 \frac{1}{q} -
        \bar{q} (b \bar{b} + d \bar{d})^2\\
\end{align*}
This implies that:
$$
\frac{ (d \bar{d} + b \bar{b})^2 ( \bar{a}^2 a^2 - q \bar{q}d^2 \bar{d}^2)}
{q d^2 \bar{d}^2 } =0 
$$
and $ |a|^2 = |q| |d|^2 $.
\begin{align*}
D_{41}  &=\underset{k}{ \sum} H_{4k} R14^{-1}_{k1} - 
        \underset{m}{ \sum} R14^*_{4m} H_{m1} \\ 
        0 &= H_{44} \frac{1}{p} - \bar{p} H_{11} \\
        0 &= (b \bar{b} + d \bar{d})^2 \frac{1}{p} -
        \bar{p} (a \bar{a} + c \bar{c})^2\\
\end{align*}
This implies that:
$$
\frac{ (d \bar{d} + b \bar{b})^2 ( p \bar{p} \bar{a}^2 a^2 - d^2 \bar{d}^2)}
{p d^2 \bar{d}^2 } =0
$$
and $ |p| |a|^2 = |d|^2 $.
This produces the third family of solutions.

\section*{Case: $ R01 $}
Consider the matrix $ R01 $.
$$
 \mathit{R01} := \begin{bmatrix}
1 & 0 & 0 & 1 \\
0 & -1 & 0 & 0 \\
0 & 0 & -1 & 0 \\
0 & 0 & 0 & 1
\end{bmatrix}
$$
The inverse and conjugate transpose of this matrix are
$$
 \mathit{R01}^{-1} = \begin{bmatrix}
1 & 0 & 0 & -1 \\
0 & -1 & 0 & 0 \\
0 & 0 & -1 & 0 \\
0 & 0 & 0 & 1
\end{bmatrix}
 \mathit{R01^*} := \begin{bmatrix}
1 & 0 & 0 & 0 \\
0 & -1 & 0 & 0 \\
0 & 0 & -1 & 0 \\
1 & 0 & 0 & 1
\end{bmatrix}.
$$
We observe 
\begin{align*}
D_{41}  &= \underset{k}{ \sum} H_{4k} R01^{-1}_{k1} - 
        \underset{m}{ \sum} R01^*_{4,m} H_{m,1} \\ 
        0 &= H_{41} - H_{11} - H_{41} \\
        0 &= H_{11}. \\
\end{align*}
If $ H_{11} =0 $ then $ Q $ is not an invertible matrix.
This matrix produces no unitary solutions.

\section*{Case: $ R02 $}
Consider the matrix $ R02 $.
$$
 \mathit{R02} := \begin{bmatrix}
1 & 0 & 0 & 1 \\
0 & 1 & 1 & 0 \\
0 & 1 & -1 & 0 \\
-1 & 0 & 0 & 1
\end{bmatrix}
$$
The inverse and conjugate transpose of this matrix are
$$
 \mathit{R02}^{-1} = \begin{bmatrix}
\frac{1}{2} & 0 & 0 & - \frac{1}{2}\\
0 & \frac{1}{2} & \frac{1}{2} & 0 \\
0 & \frac{1}{2} & - \frac{1}{2} & 0 \\
\frac{1}{2} & 0 & 0 & \frac{1}{2}
\end{bmatrix}
 \mathit{R02^*} := \begin{bmatrix}
1 & 0 & 0 & -1 \\
0 & 1 & 1 & 0 \\
0 & 1 & -1 & 0 \\
1 & 0 & 0 & 1
\end{bmatrix}.
$$
Note that $ R02 $ and $ R02^* $ is not a unitary matrix.  However, 
in \cite{hier}, the solutions are families determined by constants and 
$ Q \otimes Q $. We multiply the matrix $ R02 $ by the constant $ 2^{-\frac{1}{2}} $ and consider the modified matrix, $ R02'$.

$$
 \mathit{R02'} := \begin{bmatrix}
\frac{1}{ \sqrt{2}} & 0 & 0 & \frac{1}{ \sqrt{2}}  \\
0 & \frac{1}{ \sqrt{2}}  & \frac{1}{ \sqrt{2}}  & 0 \\
0 & \frac{1}{ \sqrt{2}}  & -\frac{1}{ \sqrt{2}}  & 0 \\
-\frac{1}{ \sqrt{2}}  & 0 & 0 & \frac{1}{ \sqrt{2}} 
\end{bmatrix}
$$
The inverse and conjugate transpose of this matrix are
$$
 \mathit{R02'}^{-1} = \begin{bmatrix}
\frac{1}{ \sqrt{2}}  & 0 & 0 & - \frac{1}{ \sqrt{2}} \\
0 & \frac{1}{ \sqrt{2}}  & \frac{1}{ \sqrt{2}}  & 0 \\
0 & \frac{1}{ \sqrt{2}}  & -\frac{1}{ \sqrt{2}}  & 0 \\
\frac{1}{ \sqrt{2}}  & 0 & 0 & \frac{1}{ \sqrt{2}} 
\end{bmatrix}
 \mathit{R02'^*} := \begin{bmatrix}
\frac{1}{ \sqrt{2}}  & 0 & 0 & -\frac{1}{ \sqrt{2}}  \\
0 & \frac{1}{ \sqrt{2}}  & \frac{1}{ \sqrt{2}}  & 0 \\
0 & \frac{1}{ \sqrt{2}}  & -\frac{1}{ \sqrt{2}}  & 0 \\
\frac{1}{ \sqrt{2}}  & 0 & 0 & \frac{1}{ \sqrt{2}} 
\end{bmatrix}.
$$
The matrix $ R02' $ is unitary, and we determine the family produced by $ R02' $.
\begin{align*}
D_{11}  &= \underset{k}{ \sum} H_{1k} R02'^{-1}_{k1} - 
        \underset{m}{ \sum} R02'^*_{1m} H_{m1} \\ 
        0 &=  H_{11} R02'^{-1}_{11} + H_{14} R02'^{-1}_{41} 
			-R02^*_{11} H_{11} - R02^*_{14} H_{41} \\
        0 &= \frac{1}{\sqrt{2}}( H_{11} + H_{14} -H_{11} + H_{41})\\
	   &=  \frac{1}{ \sqrt{2}}(H_{14}+ H_{41}). 
\end{align*}
Hence, 
\begin{equation*}
  c=-\frac{a \bar{b}}{\bar{d}}
\end{equation*}
Using this result, we determine that:   
\begin{align*}
D_{14}  &= \underset{k}{ \sum} H_{1k} R02'^{-1}_{k4} - 
        \underset{m}{ \sum} R02'^*_{1m} H_{m4} \\ 
        0 &=  H_{11} R02_{14}'^{-1} + H_{14} R02_{44}'^{-1}
		 - R02'^*_{11} H_{14} - R02'^*_{14} H_{44} \\
        0 &= -\frac{1}{ \sqrt{2}}(H_{11} - H_{44}. 
\end{align*}
This reduces to: $ |a| = |d| $. Hence, this forms 
the fourth family 
 of solutions 
to the Yang-Baxter equation.

\section*{Case: $ R03 $}
Consider the matrix $ R03 $. This is a unitary matrix and
$ R03 =R03^* = R03^{-1} $. 
$$
\mathit{R03} = \begin{bmatrix}
1 & 0 & 0 & 0 \\
0 & 0 & 1 & 0 \\
0 & 1 & 0 & 0 \\
0 & 0 & 0 & 1
\end{bmatrix}
$$
We examine the equation
\begin{align*}
D_{ij}  &= \underset{k}{ \sum} H_{ik} R03^{-1}_{kj} - 
        \underset{m}{ \sum} R03^*_{im} H_{mj} \\ 
0 &= \underset{k}{ \sum} H_{ik} R03_{kj} -  R03_{ik} H_{kj} H_{22} (-\frac{1}{pq} +1)
\end{align*}
Computations will demonstrate that $ D $ is the zero matrix. 
This case gives rise to the solutions of type 5. 

\section{Solutions to the Bracket Equation}\label{brac}

The Kauffman bracket skein relation  
determines solutions to the braided Yang-Baxter equation, \cite{knotphys}. 
We determine which 4 $ \times $ 4 matrix solutions of
 the bracket skein relation are 
unitary. These solutions are a subcase of the family 3
indicated in Theorem \ref{soln}.
If $ \hat{R} $ is a 4 $ \times $ 4 matrix 
solution to the bracket skein relation then
it satisfies the following equation. 
$$
  \hat{R} = \alpha I_4 + \alpha^{-1} U
$$
where $ U = N \odot K $ and $ N, K$ are 2 $ \times $ 2 dimensional matrices.
The operator $ \odot $ is defined for $ N $ and $ K $
\begin{align*}
  N &= \begin{bmatrix} a & b \\
    c & d \end{bmatrix} \\
  K &= \begin{bmatrix} g & h \\
    k & l \end{bmatrix} \\
  N \odot K  &= \begin{bmatrix} ag & ah & ak & al \\
    bg & bh & bk & bl  \\
    cg & ch & ck & cl  \\
    dg & dh & dk & dl  \end{bmatrix}
\end{align*}
Refering to \cite{knotphys}
the matrix $ U $ has the property that
$$
 U^2 = -( \alpha ^2 + \alpha ^{-2})U.
$$
We will denote $  -( \alpha^2 + \alpha^{-2} ) $ as $ \delta $. 
We will assume that $ \alpha $ has norm one and that
$  \alpha = e^{i \theta} $.  We obtain the following facts from 
\cite{knotphys}.
\begin{align*}
 \hat{R}^{-1} &= \alpha^{-1} I_4 + \alpha U \\
  K &= N^{-1} \\
  \delta &= \underset{a,b}{ \sum } N_{ab} N_{ab}^{-1}
\end{align*}
If $ \hat{R} $ is unitary, we obtain the
 additional restriction that $ \hat{R}^* = \hat{R}^{-1} $.
\begin{align*} 
 \hat{R}^* &= \hat{R}^{-1} \\
 \bar{ \alpha } I_4 +  \bar{ \alpha }^{-1} U^* &= \alpha^{-1} I_4
 + \alpha U \\
\end{align*}
Recall that $ \alpha $ has norm one implying that $ 
\alpha = \bar{ \alpha}^{-1}$. From this fact, we obtain $ U = U^* $. 
We determine that
$ \bar{ N } = N^{-1} $ from
\begin{align*}
  U &= U^* \\
  N \odot N^{-1} &= \bar{N}^{-1} \odot \bar{N}.
\end{align*}
We use the fact that $ \bar{N} = N^{-1} $ to determine the 
value of $ \alpha $.
Note that this argument did not refer to the dimensionality of $ N $. 
If $ | \alpha | = 1 $, then $ \hat{R} $ is a 4 $ \times $ 4 matrix. 
We demonstrate this for an n $ \times $ n matrix $ N $ 
by calculating $ \delta $ and the trace of
$ N \bar{N} $. Note that $ N \bar{N} =I_n $. 
\begin{align*} 
 \delta &= \underset{i,j}{ \sum } N_{ij}\bar{N}_{ij} \\
 \delta &= \underset{k}{ \sum}  N_{kk}\bar{N}_{kk} + 
 \underset{ i \neq j}{ \sum} N_{ij} \bar{N}_{ij} 
\end{align*}
We compute that value of the trace of $ N \bar{N} $.
\begin{align*}
  trace(N \bar{N}) = n &=  \underset{j,k}{ \sum } N_{kj}\bar{N}_{jk}\\
   n &=  \underset{k}{ \sum } N_{kk}\bar{N}_{kk} +
   \underset{ i \neq j}{ \sum} N_{ij} \bar{N}_{ji} \\
\end{align*}
Combining these two calculations, we determine that
\begin{align*}
   \delta - n =  \underset{ i \neq j}{ \sum} N_{ij} \bar{N}_{ij}-
   \underset{ i \neq j}{ \sum} N_{ij} \bar{N}_{ji} \\
   \delta -n = \underset{ i \leq j}{ \sum } | N_{ij} - N_{ji} |^2
\end{align*}
Hence, if $N $ is a n $ \times $ n matrix then 
 $ \delta \geq n $. This inequality  contradicts the fact that 
$ \delta = -( \alpha^2 + \alpha^{-2}) = -( e^{2i \theta} + e^{-2i \theta} ) $, unless $ n \leq 2 $.
We consider the specific case when $ N $ is a 2 $ \times $ 2 matrix. 
\begin{align*}
N \bar{N} &= N N^{-1} \\
N \bar{N}  &= \begin{bmatrix} a \bar{a} + b \bar{c} & \bar{b} a + b \bar{d} \\
             \bar{a} c + \bar{c} d & \bar{b} c + d \bar{d} \end{bmatrix} \\
N \bar{N} &= \begin{bmatrix} 1 & 0 \\
             0 & 1 \end{bmatrix}
\end{align*}
We examine the case when N is a 2 $ \times $ 2 matrix. Recall that
$ \delta = \underset{a,b}{ \sum} N_{a,b} N_{a,b}^{-1}$. Note that       
$
 \underset{a,b}{ \sum} N_{a,b} N_{a,b}^{-1} =
 \underset{a,b}{ \sum} N_{a,b} \bar{N}_{a,b} $, since $ \bar{N} = N^{-1} $.
From these facts we obtain
\begin{align*}
  \delta &= a \bar{a} + b \bar{b} + c \bar{c} + d \bar{d} \\
\delta    &= 1 -b \bar{c} + b \bar{b} + c \bar{c} + 1- \bar{b} c \\
  \delta &= 2 + (b -c) ( \bar{b}- \bar{c})  
\end{align*}
so that $ | \delta | \geq 2 $.
Recall that $ \alpha = e^{i \theta } $ and
$ \delta = - ( \alpha^2 + \alpha^{-2} )$, we compute that
$$ \delta  = -2 cos( \theta). $$
Implying that
$ | \delta | \leq 2 $. 
These inequalities indicate that $ | \delta | = 2$.
Using this result, we determine that
$ \alpha = i $ and $ b=c $.
With the restrictions that $ b =c $ and $ \bar{N} = N^{-1} $ then
$$
N = \begin{bmatrix} r e^{ig} & \sqrt{ 1-r^2} e^{i \frac{p}{2}} \\
  \sqrt{ 1-r^2} e^{i \frac{p}{2}} &  r e^{i(p-g)}
 \end{bmatrix}
$$
We determine the family from Theorem \ref{soln} of this solution.
We 
determine a matrix $ Q $ such that 
\begin{equation} \label{orig}
  \hat{R}_M= (Q \otimes Q) \hat{R} (Q \otimes Q)^{-1} ,
\end{equation}
 where $ \hat{R}_M $
 is a representative of a family from Theorem \ref{uybe}
 constructed from a matrix $ M $.
The family of the $ \hat{R} $ is also dependent on the value of $ Q $.

 Notice that
\begin{align*}
 \hat{R_M} &= (Q \otimes Q) \hat{R} (Q \otimes Q)^{-1} \\
\alpha Id_4 + \alpha^{-1}
 M \odot \bar{M}  &= (Q \otimes Q)\alpha I_4 + \alpha^{-1}
 N \odot \bar{N} ( Q \otimes Q)^{-1} \\
\alpha Id_4 + \alpha^{-1}
 M \odot \bar{M} &= \alpha I_4 + \alpha^{-1}(Q \otimes Q)
  N \odot \bar{N} ( Q \otimes Q)^{-1} 
\end{align*}
Note that
$$
  (Q \otimes Q) N \odot \bar{N}
(Q \otimes Q)^{-1} = (Q N Q^t) \odot ( Q^{-1} N (Q^{-1})^t ).
$$ 
The conditions of the equation \ref{orig} are equivalent to 
determining a matrix $ Q $ such that
$ Q N Q^t = M $.
If $ Q $ has the following form: 
$$ 
Q = \begin{bmatrix} x & 0 \\
  z & r^{1/2} x \end{bmatrix}
$$ 
and 
$$
 z =  \frac{-i  \sqrt{ 1-r^2} e^{ i( \frac{p}{2}-g) }}{
     \sqrt{r}}
$$ 
then 
$$
M = \begin{bmatrix}  x^2 r e^{ig} & 0 \\
  0 & x^2  e^{i(p-g)} \end{bmatrix}.
$$

We determine the family of $ \hat{R}_M $  by
considering a generic  2 $ \times $ 2 diagonal matrix.
$$
  M = \begin{bmatrix} x & 0 \\
    0 & y \end{bmatrix} \\
$$
Then
\begin{align*}
  M \odot M^{-1} &= \begin{bmatrix} 1 & 0 & 0 & \frac{x}{y} \\
    0 & 0 & 0 & 0 \\
    0 & 0 & 0 & 0 \\
    \frac{y}{x} & 0 & 0 & 1 \end{bmatrix} \\
  \hat{R_M} = \alpha I_4 + \alpha^{-1} M \odot M^{-1} &=
  \begin{bmatrix} 0 & 0 & 0 & \alpha^{-1} \frac{x}{y} \\
    0 & \alpha & 0 & 0 \\
    0 & 0 & \alpha & 0 \\
    \alpha^{-1}\frac{y}{x} & 0 & 0 & 0 \end{bmatrix}
\end{align*}
The matrix $ \hat{R_M} $ is a solution to the braided Yang-Baxter equation.
To obtain a solution to the algebraic Yang-Baxter equation, we multiply
 the matrix $ \hat{R_M} $ by the matrix $ T $ given in Theorem \ref{soln}.
$$
\hat{R_M} T  = R_M = \begin{bmatrix} 0 & 0 & 0 & \alpha^{-1} \frac{x}{y} \\
    0 & 0 &  \alpha  & 0 \\
    0 &  \alpha & 0 & 0 \\
    \alpha^{-1}\frac{y}{x} & 0 & 0 & 0 \end{bmatrix}
$$
The matrix $ \hat{R}_M $ is an element of family 2 or family 3, dependent on 
the value of $ Q $.  
 
Finally, we observe that the $ R $ matrix and $ Q $ produced satisfy the
conditions of family 3 from 
Theorem \ref{uybe}

\end{document}